\documentclass[aps,prapplied,noshowpacs,noshowkeys,amsmath,amssymb,amsfonts,reprint,longbibliography]{revtex4-1}
\usepackage{graphicx}
\usepackage{dcolumn}
\usepackage{bm}
\begin{document}
\title{Magnetic Propulsion of Self-assembled Colloidal Carpets: \\
Efficient Cargo Transport via a Conveyor Belt Effect}
\author{Fernando Martinez-Pedrero$^1$}
\author{Pietro Tierno$^{1,2}$}
\email{ptierno@ub.edu}
\affiliation{
$^1$Estructura i Constituents de la Mat\`eria, Universitat de Barcelona, 08028 Barcelona, Spain.\\
$^2$Institut de Nanoci\`encia i Nanotecnologia, IN$^2$UB, Universitat de Barcelona, Barcelona, Spain.}
\date{\today}
\begin{abstract}
We demonstrate a general method to assemble and propel highly 
maneuverable colloidal carpets which can be steered via remote control
in any direction of the plane. These colloidal micropropellers are
composed by ensemble of spinning rotors, and can be readily used to
entrap, transport and release biological cargos on command 
via an hydrodynamic conveyor-belt effect. An efficient control 
of the cargo transportation combined with remarkable "healing" 
ability to surpass obstacles, demonstrate a great potential towards 
development of multifunctional smart devices at the microscale.
\end{abstract}
\maketitle
Dynamic self-assembly
processes are widespread in physical systems,
and occur under out-of-equilibrium conditions
when for example, a driving field
supplies energy and sustains a complex and otherwise unstable
structure.~\cite{Whi02}
This powerful technique enables 
self-assembled systems adapting
their environment or performing functional and
sometimes programmable tasks. 
Examples at the microscale are disparate,
including coherent patterns from ensemble
of driven~\cite{Grz00,Sap03,Sne11,Tim13} or 
active~\cite{Wang13,But13,Sot14,Gin15}
systems.\\ 
Time-dependent magnetic fields applied to polarizable
particles represent a convenient way to assemble
and propel microscopic matter in a fluid medium,
since these fields do not affect the dispersing medium
or alter biological systems unless the latter contain
magnetic parts. The fabrication of magnetically
driven artificial propellers is a research field
of growing interest due to its direct
application in biomedicine~\cite{Sud06,Guo08,Nel10},
targeted drug delivery~\cite{Kim13}, and
microfluidics.~\cite{San11}.
In addition, practical applications
often require the use of micromachines
capable to load or unload a defined
cargo on command, or to transport the latter
in a defined place of a microfluidic platform
or a biological network.
In this context, magnetic prototypes of various form and shape
have been realized so far by using DNA
linked magnetic colloidal particles~\cite{Dre05,Tie08},
helical structures~\cite{Zha09,Fis11} or other types
of hybrid systems~\cite{Pak11,Wil14,Qiu14}.
Besides few recent examples~\cite{Sne09,Tie08}, propelling
structures based on pure dynamic self-assembly
where cooperative interactions
between the composing units
lead to coherent motion of the whole
system are still scarce.
These structures can be used to transport efficiently biochemical cargos or 
to assemble, to move and to disperse the individual units in different locations.\\
By applying suitable magnetic manipulation techniques,
we experimentally demonstrate a way to manipulate and propel
large ensemble of microscopic particles
assembled into highly ordered and
maneuverable two-dimensional carpets
via time-averaged dipolar forces.
These carpets can be assembled or disassembled
at will, rotated or transported in
any direction of the plane via
magnetic control.
\begin{figure}[t]
\begin{center}
\includegraphics[width=\columnwidth,keepaspectratio]{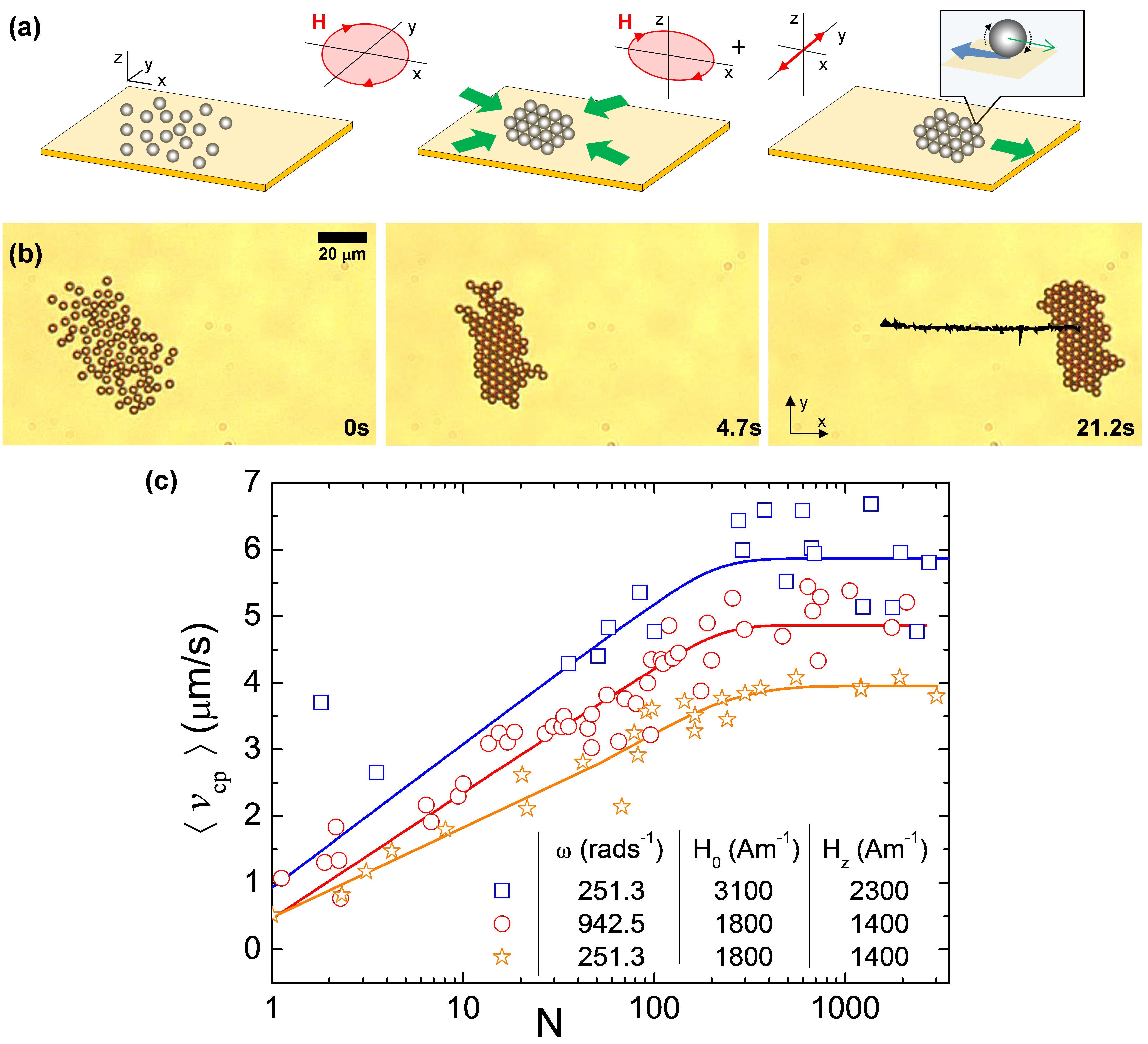}
\caption{Schematics showing the procedure to assemble
(left-middle) and propel (middle-right) a colloidal carpet. 
Assembly is induced by a magnetic field rotating in the particle 
plane $(x,y)$, whereas propulsion is induced by 
a rotating field in the perpendicular plane $(x,z)$ 
plus an oscillating component along the $y-$axis. (b)
Snapshots of a colloidal carpet assembled and 
propelled by an applied field with amplitudes
$H_0= 1800A m^{-1}$, $H_z=1400A m^{-1}$ and frequencies $\omega=942.5 rad s^{-1}$,
$\omega_y=471.2rad s^{-1}$ (MovieS1 in~\cite{EPAPS}).
(c) Semilog plot of the average speed $\langle v_{cp} \rangle$ 
versus number of particles
$N$ composing the carpet at 
different field values and frequencies 
(here, $\omega_y=\omega/2$). }
\label{fig_1}
\end{center}
\end{figure}
Further we show the possibility to entrap,
transport and release micro-objects
such as biological cells or colloidal cargos
in a fluidic environment, by using the hydrodynamic
conveyor-belt generated by the moving structure.
These magnetic carpets thus represent an alternative
class of propelling prototypes based on dynamic self assembly.\\
As building blocks for the colloidal carpets
we used $2.8 \mu m$ size monodisperse 
paramagnetic colloids (Dynabeads M-270, Invitrogen)
composed by a polymer matrix 
and evenly doped with
superparamagnetic grains ($\sim 14 \%$ iron oxide 
content, density
$\rho=1.3 g cm^{-3}$).
The original aqueous suspension of the particles
($\sim 7\cdot 10^9$ beads/ml) was diluted with high deionized water
($18.2 M\Omega \cdot cm$, MilliQ system) at a concentration of $\sim 1.4\cdot 10^8$ beads/ml. To prevent convection
effects,
the colloidal
suspension is sandwiched between a glass slide (Corning Incorporated)
and a microscope coverslip
(Agar Scientific), both separated by a double faced adhesive
tape. As colloidal cargos we use
either budding yeast cells ({\it Saccharomyces
Cerevisiae}) or silica dioxide microspheres
($5 \mu m$ size, density $\rho = 2.1 g cm^{-3}$)
obtained from Sigma Aldrich.\\
The measurement cell is placed in the center
of two orthogonal pairs of coils arranged on the
stage of a light microscope (Eclipse Ni, Nikon),
and aligned along
the $y$ and $x$ axis. A fifth coil aligned along $z$ is
centered just under the sample cell. To apply
time-dependent magnetic fields the coils
are connected with a waveform generator (TGA1244, TTi)
feeding power amplifiers (AMP-1800, AKIYAMA and BOP
20-10M, Kepco).\\
The paramagnetic colloids can be easily
magnetized by a relatively low external field
$\bm{H}$, acquiring a dipole moment $\bm{m}=V\chi \bm{H}$ pointing along
the field direction. Here $V$ is the particle volume and 
$\chi=0.4$ the magnetic susceptibility under a static field.
For a rotating field circularly polarized 
in the $(x,z)$ plane,
$\bm{H}=H_0(\cos{(\omega t)}\bm{e}_x-\sin{(\omega t)}\bm{e}_z)$,
the particle dynamics becomes more complicated due to
the presence of a finite magnetization relaxation
time $\tau_{r}$~\cite{Tie07,Jan09}.
The average magnetic torque applied to the particle
can be calculated as
$\bm{T}_{m}=\mu_0 \langle \bm{m} \times \bm{H} \rangle$,
where $\mu_0 = 4 \pi \cdot 10^{-7} \, H m^{-1}$
and $\langle ... \rangle$ denotes a time average.
Solving the relaxation equation for
the magnetic moment~\cite{Ceb11}
we find that,
$\bm{T}_{m}=\frac{\mu_0 V \chi H_0^2 \tau_{r} \omega}{1+\tau_{r}^2\omega^2}\bm{e}_y$,
where $\tau_r\sim 10^{-4}s$ from independent measurements (data not shown).
The applied torque forces the particle to rotate
at an average angular velocity $\bm{\Omega}$
in the fluid medium.
Upon balancing ${\bm T}_m$ with
the viscous torque
arising from the rotation in the medium 
${\bm T}_v=-8\pi \eta {\bm \Omega}a^3$
the average rotational speed reads as
$\langle \Omega \rangle = \mu_0 H_0^2\chi \tau_{r} \omega /6\eta (1+ \tau_{r}^2 \omega^2)$.
Here $\eta=10^{-3} Pa \cdot s$ denotes the dynamic viscosity of water.\\
Pair of rotors 
also interact via 
dipolar forces. 
The interaction potential
between two equal dipoles
$(i,j)$, at a distance
$\bm{r}$ away,
is given by
$U_{dd}=\frac{\mu_0}{4\pi}(\frac{{\bm m}_i {\bm m}_j}{r^3}-\frac{3({\bm m}_i \cdot {\bm r})({\bm m}_j\cdot {\bm r}) }{r^5})$,
and is maximally attractive (repulsive) for particles with
magnetic moments parallel (normal) to $\bm{r}$.
Time averaging $U_{dd}$
for a rotating magnetic field in the $(x,y)$
plane gives an  effective attractive potential
in this plane $\langle U_{dd} \rangle =-\frac{\mu_0 m^2}{8 \pi (x+y)^3}$.
This interaction potential
can be thus used to magnetically
assemble highly ordered 2D particle monolayers~\cite{Ost09,Yan14}.\\
Fig.1(a) shows the general procedure
to realize and to propel a magnetic carpet.
First compact clusters are realized by applying
an external uniform magnetic field circularly polarized 
in the $(x,y)$ plane parallel to the glass surface, 
$\bm{H}\equiv H_0(\cos{(\omega t)}\bm{e}_x-\sin{(\omega t)}\bm{e}_y)$,
being  $\omega$ the angular frequency and $H_0$ the amplitude.
The applied field induces dipolar interactions
between the magnetic colloids which are in 
average attractive, and the particles rapidly 
assemble into a close packed cluster free
of defects or vacancies. The colloidal clusters 
are stable only in presence of the driving field, 
while immediately disintegrate due to thermal 
fluctuations once the applied field is switched 
off. As reported in previous works on assembly of 
"magnetic holes"~\cite{Hel08} or Janus colloids~\cite{Yan15}, 
we observe that when the cluster is formed, 
it continues to rotate but at a smaller 
angular frequency than that imposed by the 
driving field. The cluster rotation arises 
from an unbalanced viscous shear force 
experienced by the torque particles located 
at the edge of the cluster~\cite{Yan15,Sch12}.
Once the carpets are formed, propulsion is obtained by 
applying a rotating field in the ($x,z$) plane plus an
additional component oscillating with angular 
frequency $\omega_y$
along the perpendicular direction($y$). 
The total field is given by:
\begin{equation}
\bm{H}\equiv H_0 (\cos{(\omega t)}\bm{e}_x+\sin{(\omega_y t)}\bm{e}_y-\frac{H_z}{H_0}\sin{(\omega t)}\bm{e}_z),
\end{equation}
as depicted in the right part of Fig.1(a).
In most of the experiments reported here we used 
$H_0 \equiv H_x = H_y$ and $\omega_y = \frac{\omega}{2}$.
%
\begin{figure}[t]
\begin{center}
\includegraphics[width=\columnwidth]{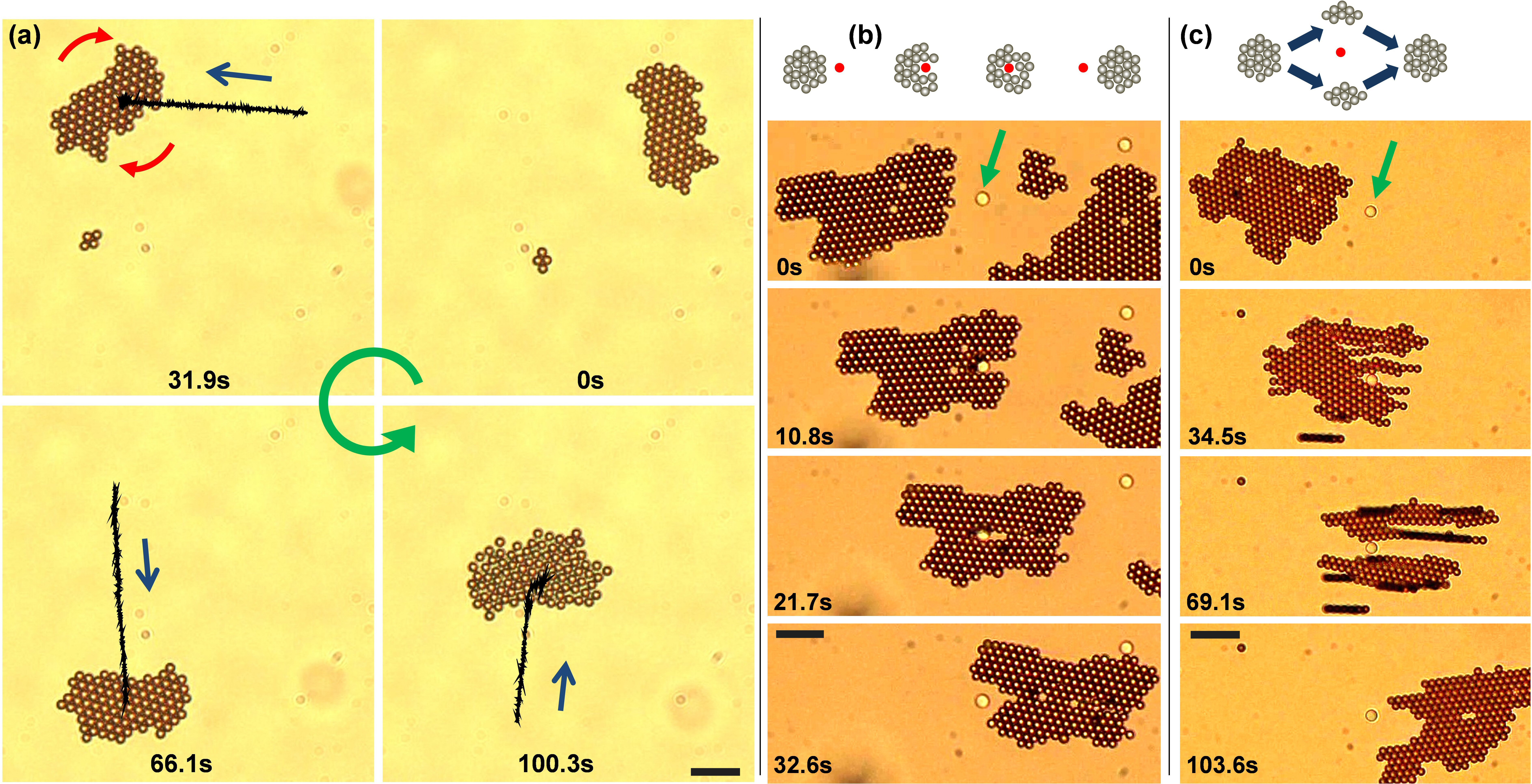}
\caption{(Color online)(a) Snapshots in counter clockwise order, 
showing the guided motion of a colloidal carpet under dynamic magnetic fields 
with $H_0= 2300A m^{-1}$, $H_z=1400A m^{-1}$ 
and $\omega=942.5 rad s^{-1}$, $\omega_y= \omega/2$ (MovieS2a).
Center of mass trajectory is superimposed as a black line. 
(b and c) Top: schematic, bottom: sequence of images of colloidal
carpets moving against a $5 \mu m$ immobile silica particle. 
In (b) the field parameters remained 
constant ($H_z= 1400 A m^{-1}$, $H_0 =1800 A m^{-1}$ and frequencies 
$\omega = 942.5 rad s^{-1} =2\omega_y$, MovieS2b) while in (c) 
the field $H_y$ was set to zero and
$H_z$ decreased to $H_z= 560 A m^{-1}$ after $t=26s$ (MovieS2c).
Later ($t=92.8s$) the $H_y$ field is restored and the carpet forms again.
The scale bars in all images are $20 \mu m$.}
\label{fig_2}
\end{center}
\end{figure}
%
The rotating field in the $(x,z)$ plane exerts
a torque on the individual particles
forcing them to rotate close
to the glass substrate.
The particles acquire a
net translational motion with an average speed
$\langle v_x \rangle \sim  \Omega a$,
due to the hydrodynamic coupling with the substrate~\cite{Gol67},
making possible the propulsion of the whole carpets.
The $H_y$ component helps to keep the structure
compact during motion, avoiding lateral
separation of the constituent particles.
The choice of $\omega_y = \omega/2$ avoids the carpet
rotation, which otherwise would occur by simply
setting $\omega_y = \omega$. Fig.1(b) shows a carpet propelling
with an average speed of $\langle v_{cp} \rangle = 2.7 \mu m s^{-1}$
along the $x$ direction, keeping its structure
intact. During motion the carpet
trajectory is quite stable, presenting negligible
displacements in the perpendicular direction.
We note that in contrast to the assembly stage,
the rotating field
in the $(x,z)$ plane is now strongly elliptically
polarized ($H_0>>H_z$) in order to maintain the
structure confined to two dimensions.\\
Although the speed acquired by an individual
particle (rotor) moving close to the substrate is 
relatively slow, for an ensemble of rotors forming the 
carpet the effect becomes cooperative 
resulting in a faster translational motion. 
In Fig.1(c) this effect is reflected by measuring
the average propulsion speed $\langle v_{cp} \rangle$ 
as a function of the number of rotors $N$ for a series 
of carpets having approximately a spherical shape. 
The continuous dashed lines are averaged curves
from the experimental data showing the general trend. 
While $\langle v_{cp} \rangle$ initially rapidly increases with $N$,
it saturates around $N \sim 300$, corresponding to a carpet 
area $S \sim 1800 \mu m^2$. Beyond this
value, the rotors composing the colloidal structure start to 
be far away and the cooperative effect reaches its 
maximum efficiency.
%
\begin{figure}[t]
\begin{center}
\includegraphics[width=\columnwidth,keepaspectratio]{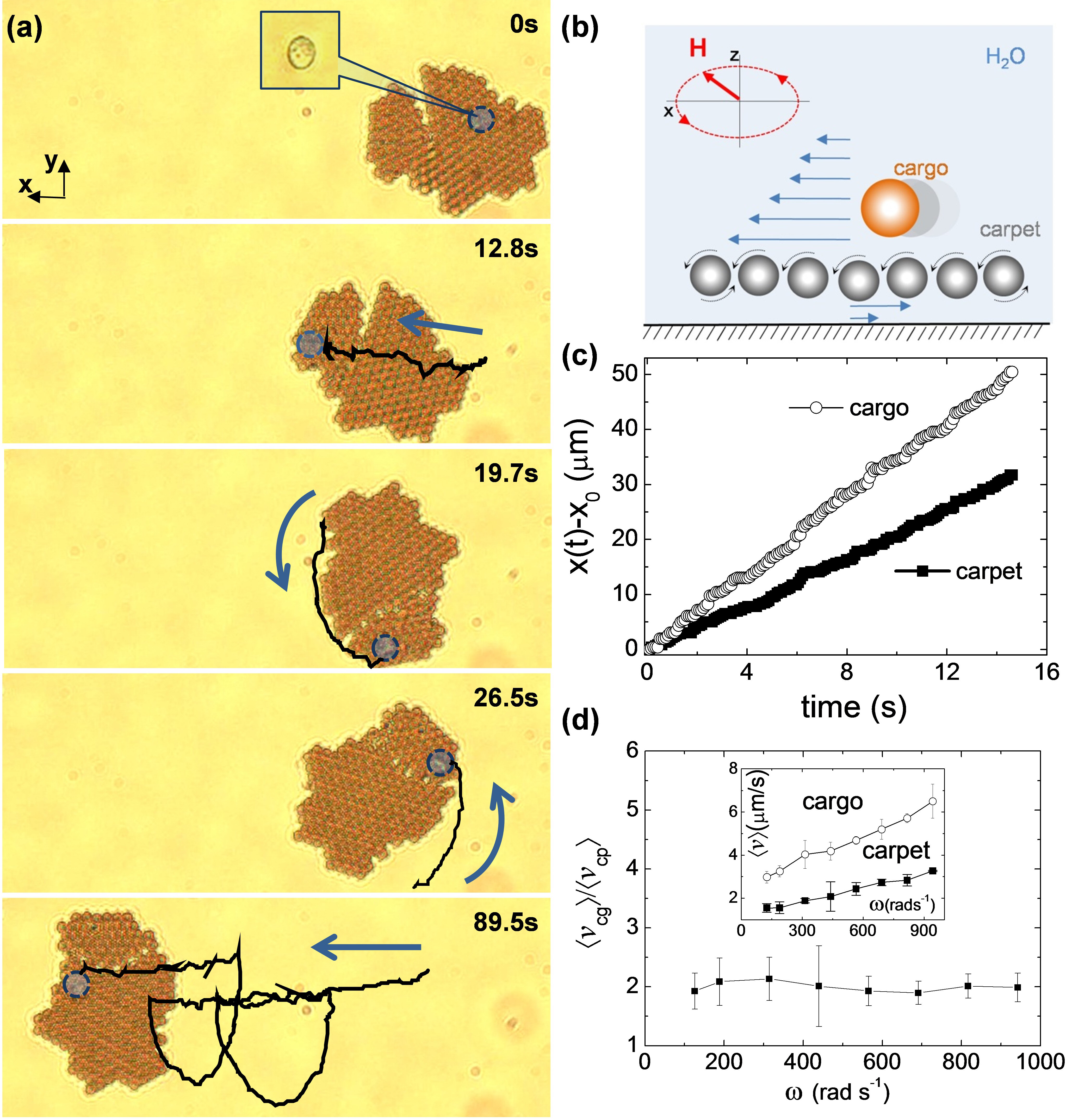}
\caption{(Color online) (a) Sequence of images showing 
the transport one yeast cell by combining 
translation and rotation of the carpet (MovieS3a). 
The position of the tracked cell is superimposed as 
a black line. The applied field parameters used 
to translate the carpet is the same as in Fig.2a.
(b) Schematic of the hydrodynamic conveyor belt 
generated by the carpet to transport a colloidal cargo. 
(c) Plots of the distance versus time for the carpet (filled squares) 
and cargo (empty circles) for Fig3(a) from $t=26.5s$ to $t=41.0s$. 
(d) Average speed of a colloidal cargo 
$\langle v_{cg} \rangle$ scaled with the carpet
speed $\langle v_{cp} \rangle$ as a function of
$\omega$ ($H_z= 1400 A m^{-1}$, $H_0 = 1800 A m^{-1}$ and 
$\omega = 2\omega_y$).
Inset shows $\langle v_{cp} \rangle$ and 
$\langle v_{cg} \rangle$ separately.}
\label{fig_3}
\end{center}
\end{figure}
%
Further, we observe that
at parity of frequency, increasing the field 
amplitude $H_0$ (orange stars and blue squares in Fig.1(c)) raises 
the average speed since the magnetic 
particles are subjected to higher magnetic 
torque and acquire a faster rotational 
motion. On the other hand, the propulsion 
speed can be also increased via the driving 
frequency (orange stars and red circles in Fig.1(c)), although this effect is less pronounced.\\
The maneuverability of our colloidal
structures is shown in Fig.2(a), where the
carpet is dynamically guided respectively towards
the left, south and north of the observation area.
As shown in MovieS2a, these turns are achieved by
temporary stopping the propulsion ($H_z=0$) and inducing
a rotational motion of the carpet by setting $\omega_y=\omega=942.5 rad s^{-1}$.
The change in direction is then obtained by restoring
the field but exchanging the phases and
frequencies of $H_x$ and $H_y$. In Figs2(b,c)
we explore the stability of the propelling
carpets by forcing them to move against an
immobile obstacle. As obstacle, we use
a $5 \mu m$ silica particle which was
permanently attached to the substrate.
Fig.2b (MovieS2b) shows the carpet dynamics
without changing the applied field. In this
case the obstacle locally melts the moving
crystal of rotors, but strong attractive 
dipolar interactions prevent its disaggregation, 
favoring re-crystallization templated by the ordered
particles surrounding the melted region. 
An alternative method to pass the obstacle 
is presented in Fig.2c (MovieS2c), which 
could be adopted to surpass larger obstacles. 
In this particular situation, we split the carpet 
into several pieces by setting temporary $H_y=0$ and
decreasing the perpendicular field $H_z$. 
The absence of the $H_y$ component makes 
the structure less stable during its 
propulsion, inducing the lateral 
separation of the constituent 
particles preferentially along 
the crystalline axis as triggered by the
obstacle. After crossing the obstacle, 
all pieces can be reassembled back by 
restoring the field parameters. We note 
that the splitting of the carpet into pieces 
leads to a final different colloidal structure.
%
\begin{figure*}[t]
\begin{center}
\includegraphics[width=0.75\textwidth,keepaspectratio]{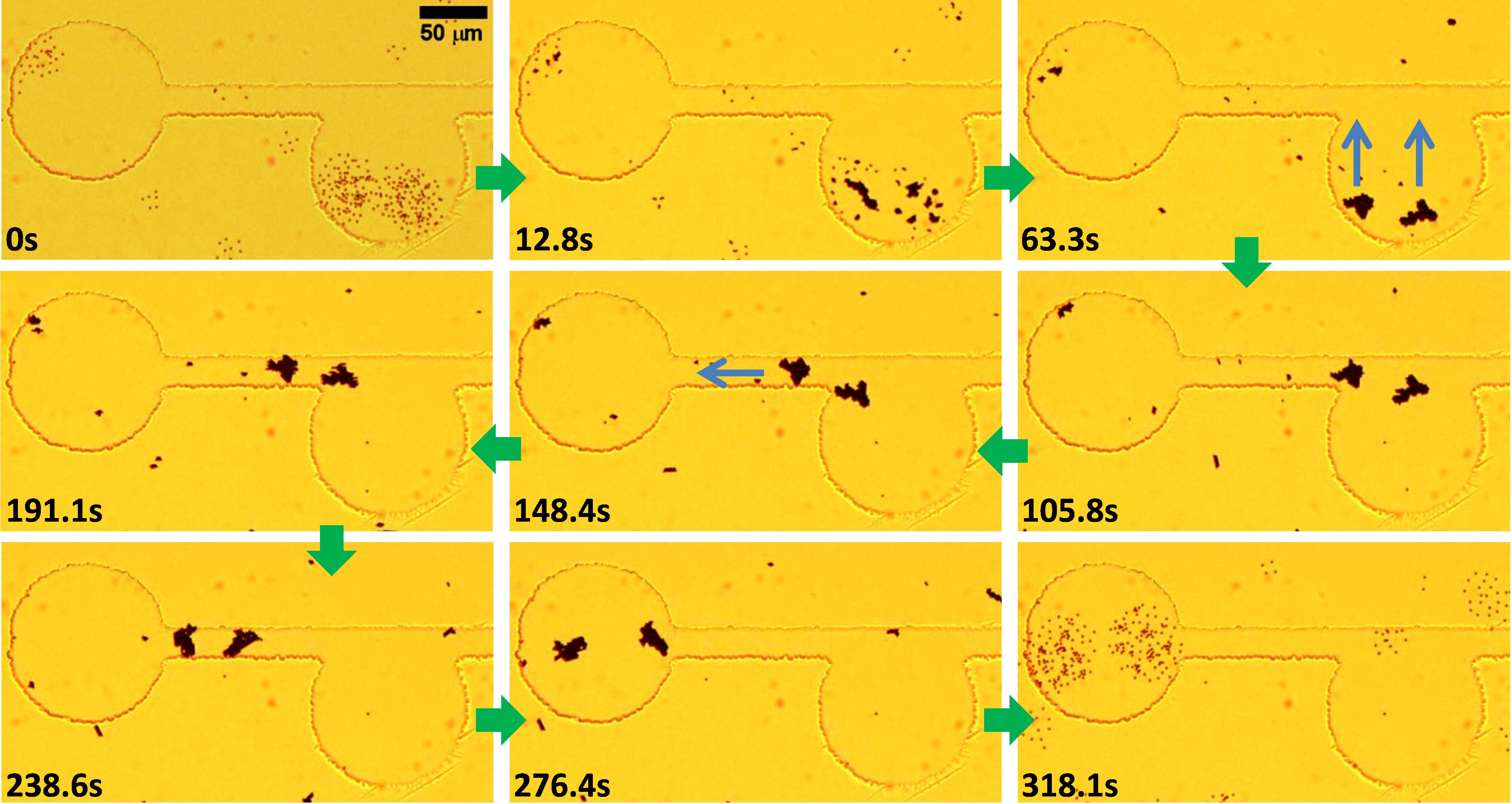}
\caption{(Color online) Sequence of images showing the
formation of two colloidal carpets
from dispersed particles (top sequence)
within a glass etched circular chamber ($115 \mu m$ diameter).
The carpets are later transported across
thin microchannels $20 \mu m$
width having (middle sequence) and finally the
particles are redispersed by switching off the applied
field $318.1s$ later (MovieS4).
The applied field parameters during transport
are $H_0= 1900 A m^{-1}$, $H_z= 980 A m^{-1}$,
$\omega_z=942.5 Hz$.
The frequencies are $\omega_x =\omega_z/2$ and
$\omega_y =\omega_z$ ($\omega_x =\omega_z$ and $\omega_y =\omega_z/2$)
when the carpets move upward (leftward).}
\label{fig_4}
\end{center}
\end{figure*}
%
Nevertheless we find that in both cases (Fig.2b and Fig.2c) 
the carpet is able to heal small wounds and to 
adapt its shape to the most compact structure 
after crossing the obstacle.\\
Non invasive approaches based on propulsive
colloids to move and displace micro-objects
have been achieved in the past by using elongated 
nanorods~\cite{Pet12}, anisotropic particles~\cite{Pal13},
helical~\cite{Tot12} or
other composite structures~\cite{Sun08,Sak10}. Our
self-assembled carpet allows to
manipulate and transport microscopic
cargos over its extended area without
requirement of direct contact or chemical
binding. This feature is demonstrated in
Fig.3(a), corresponding MovieS3a, where
a carpet having an area of $S \sim 1200 \mu m^2$ is used
to transport one yeast cell along a $95 \mu m$ straight
path. First, as shown in the Supporting videos (MovieS3b, MovieS3c),
the loading of the cargo, here a biological cell, can be realized 
by direct
entrapment (MovieS3b) 
or by
lifting the cargo when the carpet moves/rotates
close to it (MovieS3c). Once
above the magnetic carpet, the microscopic cargo
is transported by using the hydrodynamic flow
generated by the rotating particles.
As illustrated by the schematic of Fig.3(b),
this flow advects the colloidal carpet acting
as a conveyor belt. Surprisingly, in contrast
to other techniques to displace micro-objects,
we find that the cargo moves faster than the
underlying structure. For instance, in the
specific case of Fig.3(a), the yeast cell
travels at an average speed
of $\langle v_{cg} \rangle = 3.5 \mu m s^{-1}$,
twice larger than the speed of the carpet. 
This behavior is independent of the cargo nature,
as shown in Fig.3(d) where we found approximately
the same ratio by employing $5 \mu m$ silica particles.
At parity of field strength, we observe that
both $\langle v_{cg} \rangle$ and
$\langle v_{cp} \rangle$ increases linearly
with the frequency, and their ratio keeps constant.
Increasing $\omega$ increases the rotational speed of the
particles and thus the hydrodynamic flux above and
below the carpet. While the colloidal cargo is free
to be dragged by this flux, the colloidal carpet is
slower since being closer to the surface it experience
a larger hydrodynamic drag~\cite{Rus89}. Faster speed of the cargo
also implies that the latter will reach the
edge and leave the carpet in a finite period
of time. In order to transport the cargo for
a longer distance, we manipulate the carpet
inducing a $\pi/2$ rotation each time the cell
reaches the edge of the structure. In the
particular case of Fig.3(a), two of such
rotations are required to allow the yeast
cell to cover the whole observation area
in $89.5s$.
The ability to assemble and transport
colloidal carpets close to a confining
surface creates many opportunities to
integrate them in a microfluidic platform.
As an illustrative example, in
Fig.4 we assemble and guide
two colloidal carpets in a
microfluidic environment.
We realized a $4 \mu m$ depth
channel structure above a
glass plate by using a wet
etching (HF) technique. The
structure is characterized
by a network of circular
compartments having $115 \mu m$
diameters and connected by
channels $20 \mu m$ wide.
Free paramagnetic colloids
previously collected in one circular
compartment are assembled into two separate
clusters having sizes of $670 \mu m^2$ and $430 \mu m^2$ by an
external rotating field. After $t=63.3s$, the rotating
field in the perpendicular plane is used to propel and
to guide these colloidal carpets along the path connecting
the two chambers. Once reached the second compartment ($t=276.4s$),
switching off the applied field melts the carpets, dispersing the
particles evenly within the chamber. Even if no cargo has been
transported during this operation, one can use the paramagnetic
colloids directly as drug delivery vectors, since their
surface can be easily functionalized with chemical agents
in order to bind and to transport functional molecules~\cite{Yao11}.
Besides the basic geometries shown in Fig.4, circuits having
more complex patterns can be equally explored by the propelling
carpets by simply adjusting on the fly, the orientation and
direction of the driving field.\\
In conclusion we developed a simple and
versatile technique to magnetically assemble
and propel highly reconfigurable colloidal
carpets in all direction of the plane. Although
we demonstrate this method with commercial available
paramagnetic colloids, it can be easily extended
to other types of recently developed particles with
heterogeneous magnetic properties such as cubes~\cite{Sac12},
ellipsoids~\cite{Gu11}, Janus~\cite{Sin11} or
anisotropic~\cite{Zer08} ones. As
opposed to existing microscopic engines chemically powered
or magnetically propelled, the mechanism of motion of our
carpets is cooperative and based on the rectification of
the hydrodynamic flow generated by each rotor close to the
bounding wall. On the application side, we showed
that these mobile colloidal sheets can be used to
transport biological or colloidal cargos entrapped
and translated via a hydrodynamic conveyor belt
effect. Finally the carpets can be also used to
assemble, maneuver and disperse microscopic
particles through a microfluidics network,
making them suitable for fluid based microdevices.\\

We thank Ignacio Pagonabarraga Mora
for stimulating discussions. F.M.P. and P.T.
acknowledge support from the European Research
Council Project No. 335040. P.T. acknowledges 
support from "Ramon y Cajal" Program 
No.RYC-2011-07605, from Mineco (FIS2013-41144-P) 
and AGAUR (2014SGR878).
\bibliography{biblio}
\end{document}